\documentclass[12pt]{article}
\usepackage{wrapfig}
\usepackage{epsfig}
\usepackage{hyperref}
\usepackage{amssymb}
\usepackage{amsmath}
\usepackage{amsfonts}
\usepackage{latexsym}
\usepackage{wasysym}
\usepackage{multirow}
\usepackage{fixmath}
\usepackage{color}
\usepackage{stackrel}
\usepackage{txfonts}                                                            
\usepackage{bbm}

\newcommand{\bea}{\begin{eqnarray}}
\newcommand{\eea}{\end{eqnarray}}
\newcommand{\bite}{\begin{itemize}}
\newcommand{\eite}{\end{itemize}}

\newcommand{\gev}{\,{\rm GeV}}
\newcommand{\MSbar}{$\overline{\rm MS}$ }

\begin{document}

\title{
\centering{\Large \bf Isospin splittings of meson and baryon masses from three-flavor lattice QCD + QED}\\[0.25em]} 
\author{R.~Horsley$^a$, Y.~Nakamura$^b$, H.~Perlt$^c$, D.~Pleiter$^d$, P.~E.~L.~Rakow$^e$,\\ G.~Schierholz$^f$, A.~Schiller$^c$, R.~Stokes$^g$, H.~St\"uben$^h$,\\ R.~D.~Young$^g$ and J.~M.~Zanotti$^g$\\[1em] 
$^a$ School of Physics and Astronomy, University of Edinburgh,\\ Edinburgh
EH9 3FD, United Kingdom\\[0.15em] 
$^b$ RIKEN Advanced Institute for Computational Science,\\ Kobe, Hyogo 650-0047, Japan\\[0.15em] 
$^c$ Institut f\"ur Theoretische Physik, Universit\"at Leipzig,\\ 04103
Leipzig, Germany\\[0.15em]  
$^d$ J\"ulich Supercomputer Center, Forschungszentrum J\"ulich,\\
52425 J\"ulich, Germany\\[0.15em]
$^e$ Theoretical Physics Division, Department of Mathematical Sciences,\\
University of Liverpool, Liverpool L69 3BX, United Kingdom\\[0.15em] 
$^f$ Deutsches Elektronen-Synchrotron DESY,\\ 22603 Hamburg, Germany\\[0.15em] 
$^g$ CSSM, Department of Physics, University of Adelaide,\\ Adelaide SA 5005, Australia\\[0.15em]
$^h$ RRZ, University of Hamburg, 20146 Hamburg, Germany}

\date{}

\maketitle

\begin{abstract}
  Lattice QCD simulations are now reaching a precision where isospin
  breaking effects become important. Previously, we have developed a
  program to systematically investigate the pattern of flavor symmetry
  beaking within QCD and successfully applied it to meson and baryon
  masses involving up, down and strange quarks. In this Letter we
  extend the calculations to QCD + QED and present our first results
  on isospin splittings in the pseudoscalar meson and baryon
  octets. In particular, we obtain the nucleon mass difference of
  $M_n-M_p=1.35(18)(8)\,\mbox{MeV}$ and the electromagnetic
  contribution to the pion splitting
  $M_{\pi^+}-M_{\pi^0}=4.60(20)\,\mbox{MeV}$.  Further we report
  first determination of the separation between strong and
  electromagnetic contributions in the \MSbar scheme.
\end{abstract}

\clearpage
\section{Introduction and general strategy}

Isospin breaking effects are crucial for the existence of our
Universe. Our Universe would not exist in the present form if the $n -
p$ mass difference would only be slightly different. If it would be
larger than the binding energy of the deuteron, no fusion would take
place. If it would be a little smaller, all hydrogen would have been
burned to helium. Isospin breaking in hadron masses has two sources, the mass
difference of up and down quarks, and electromagnetic
interactions. Both effects are of the same order of magnitude and
cannot be separated unambiguously due to the nonperturbative nature of
the strong interactions. This makes a direct calculation from QCD +
QED necessary~\cite{QED,Horsley:2013qka,Borsanyi:2014jba}. While
substantial progress has been made, \cite{Borsanyi:2014jba} is
the only other published work to report simulations with
fully dynamical QCD + QED.

In~\cite{Bietenholz:2010jr,Bietenholz:2011qq} we have outlined a program to systematically investigate the pattern of flavor symmetry breaking in three-flavor lattice QCD for Wilson-type fermions. Our strategy was to start from the SU(3) symmetric point with all three quark masses equal, $m_u=m_d=m_s$, and extrapolate towards the physical point keeping the average sea quark mass $\bar{m}=\left(m_u+m_d+m_s\right)/3$
constant. For this trajectory to reach the physical quark masses, $\bar{m}$ is tuned to the physical value of the average pseudoscalar meson mass $X_\pi^2 = \left(M_{K^0}^2+M_{K^+}^2+2M_{\pi^0}^2-M_{\pi^+}^2\right)/3$. We denote the distance from $\bar{m}$ by
$\delta m_q = m_q-\bar{m}\ (q=u,d,s) \,.$
This implies $\delta m_u + \delta m_d + \delta m_s =0$ on our quark mass trajectory. To describe how physical quantities depend on the quark masses, we Taylor expand about the symmetric point~\cite{Bietenholz:2011qq}. This results in polynomials in $\bar{m}$ and $\delta m_q$, which we classify into representations of the SU(3) and $\mathrm{S_3}$ flavor groups. As we  keep $\bar{m}$ constant and change only the octet part of the mass matrix, to first order in $\delta m_q$ flavor symmetry is broken by an SU(3) octet, leading to Gell-Mann--Okubo mass relations. We follow a similar approach here with QED added~\cite{Horsley:2013qka}.

The symmetry of the electromagnetic current is similar to the symmetry of the quark mass matrix. The simplifications that come from $\delta m_u + \delta m_d + \delta m_s = 0$ in the mass case are analogous to the simplifications we get from the identity $e_u + e_d + e_s = 0$. 
A difference between quark mass and electromagnetic expansions is that in the mass expansion we can have both odd and even powers of $\delta m_q$, whereas only even powers of the quark charges $e_q$ are allowed. 
We consider contributions of $O(e_q^2)$ only. Hence, QED corrections can be simply read off from the mass expansion presented in~\cite{Bietenholz:2011qq}, dropping the linear terms and changing masses to charges.

\begin{figure}[!t]
\vspace*{-1cm}
   \begin{center} 
      \epsfig{file=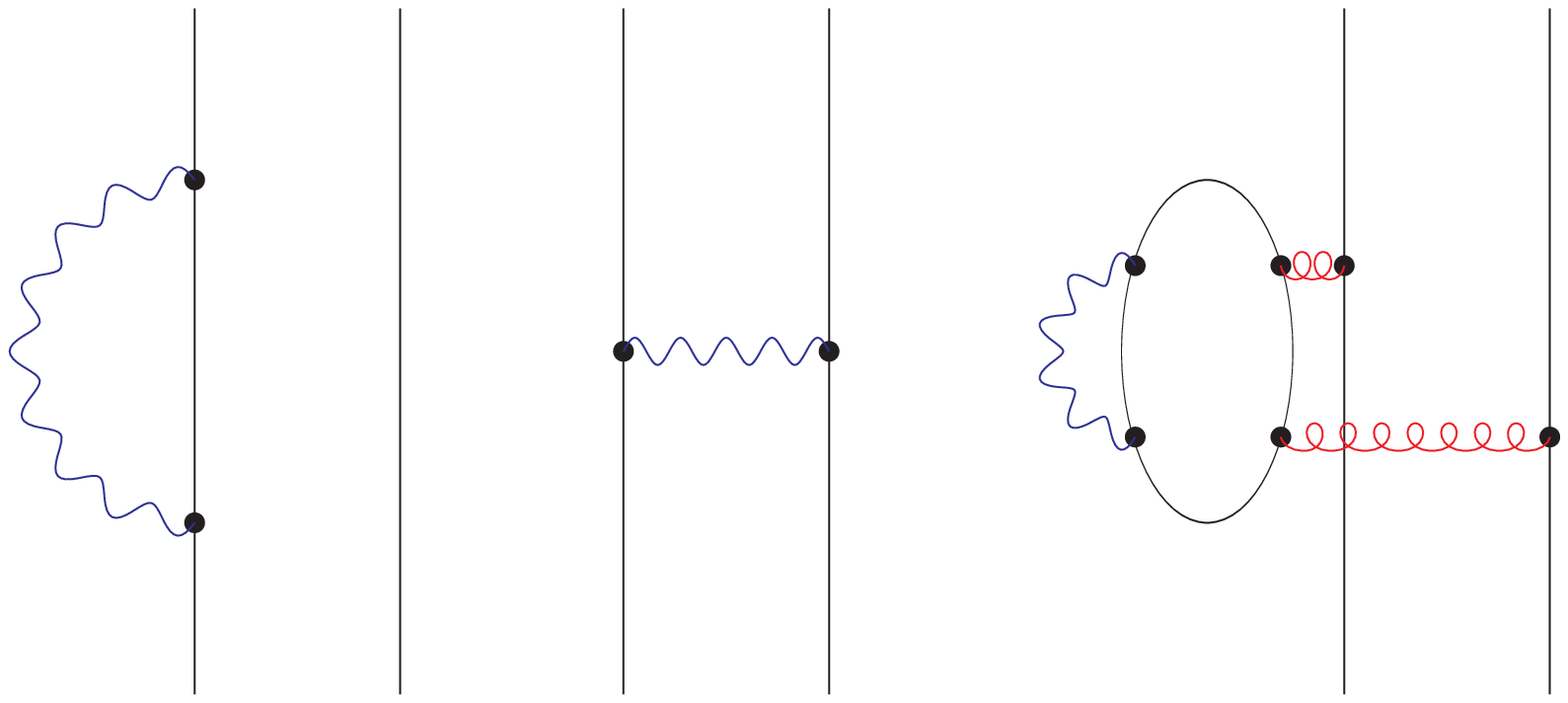,width=5cm,angle=90,clip}
   \end{center} 
\vspace*{-1cm}
\caption{Examples of Feynman diagrams contributing to the meson
  electromagnetic mass to $O(e_q^2)$. Wavy lines are photons, curly
  lines are gluons.}
\label{diag}
\end{figure}

For the masses of octet mesons with the flavor structure $a\bar{b}$,
and all annihilation diagrams turned off, we find to leading order in $\alpha_{\rm EM}$
\begin{equation}
\begin{split}
M^2(a\bar{b}) &= M_0^2 + \alpha\, (\delta m_a +\delta m_b) 
+ \beta_0^{\rm EM}\, (e_u^2+e_d^2+e_s^2) + \beta_1^{\rm EM}\,(e_a^2+e_b^2) + \beta_2^{\rm EM}\,(e_a-e_b)^2  \\
&+ \gamma_0^{\rm EM}\,(e_u^2\delta m_u+e_d^2\delta m_d+e_s^2\delta m_s)
+\gamma_1^{\rm EM}\,(e_a^2\delta m_a+e_b^2\delta m_b) \\
&+ \gamma_2^{\rm EM}\, (e_a-e_b)^2\, (\delta m_a+\delta m_b)
+ \gamma_3^{\rm EM}\, (e_a^2-e_b^2)\, (\delta m_a-\delta m_b) \\
&+ \gamma_4^{\rm EM}\,(e_u^2+e_d^2+e_s^2)\,(\delta m_a +\delta m_b) 
+ \gamma_5^{\rm EM}\,(e_a+e_b)\,(e_u\delta m_u+e_d\delta m_d+e_s\delta m_s)  
\end{split}
\label{mmass}
\end{equation}
up to corrections of $O(\delta m_q^2)$. Several of the coefficients in (\ref{mmass}) can be matched up with
different classes of Feynman diagrams shown in Fig.~\ref{diag}. 
The first diagram, with both ends of the photon attached to the same
valence quark, contributes to $(\beta_1^{\rm EM}+\beta_2^{\rm
  EM})$. 
The second diagram, with the photon crossing between the valence
lines, contributes to $\beta_2^{\rm EM}$. 
The last diagram, with the photon being attached to the sea quarks, is
an example of a diagram contributing to $\beta_0^{\rm EM}$. 
It would be missed out if the electromagnetic field was quenched
instead of dynamical. 
Similar assignments hold for the mixed (charge squared times mass)
terms.
For a single choice of sea quark masses, the $\beta_0^{\rm EM}$ and
$\gamma_4^{\rm EM}$ terms can be absorbed into the constant $M_0^2$
and the $\alpha$ term. 
However, for a combined fit of both QCD and QCD + QED data we will
need these coefficients.
More details can be found in \cite{Paul}.
Similarly, for octet baryons with the flavor structure $aab$ we find to leading order in $\alpha_{\rm EM}$
\begin{equation}
\begin{split}
M^2(aab) &= M_0^2 + \alpha_1\, (2\delta m_a +\delta m_b) + \alpha_2\,
(\delta m_a -\delta m_b)\\ 
&+ \beta_0^{\rm EM}\, (e_u^2+e_d^2+e_s^2) + \beta_1^{\rm
  EM}\,(2e_a^2+e_b^2) + \beta_2^{\rm EM}\,(e_a-e_b)^2  + \beta_3^{\rm
  EM}\,(e_a^2-e_b^2) 
\end{split}
\label{bmass}
\end{equation}
up to corrections of $O(\delta m_q^2)$. This excludes the case of baryons with three different quarks, as in
the $\Sigma^0 - \Lambda$ system~\cite{Horsley:2014koa}. 
Again, the $\beta_0^{\rm EM}$ term can be absorbed into the mass term
$M_0^2$. The coefficients $\beta_1^{\rm EM}$, $\beta_2^{\rm EM}$ and $\beta_3^{\rm EM}$ can be matched up with distinct classes of Feynman diagrams similar to the ones in Fig.~\ref{diag}.

Our goal is to compute the mass splittings of pseudoscalar mesons and
octet baryons at the physical point for QCD + QED. 
This amounts to determining the coefficients $\alpha$, $\beta^{\rm
  EM}$ and $\gamma^{\rm EM}$ in (\ref{mmass}) and (\ref{bmass}). 
It greatly helps to vary valence and sea quark masses
independently~\cite{Bietenholz:2011qq}, which is referred to as
partial quenching (PQ). 
In this case the sea quark masses remain constrained by
$\bar{m}=\mbox{constant}$, while the valence quark masses $\mu_u$,
$\mu_d$ and $\mu_s$ are unconstrained. 
Defining $\delta \mu_q = \mu_q - \bar{m}$, the resulting modification of Eq.~(\ref{mmass}) to PQ octet mesons is
\begin{equation}
\begin{split}
M^2(a\bar{b}) &= M_0^2 + \alpha\, (\delta \mu_a +\delta \mu_b) 
+ \beta_0^{\rm EM}\, (e_u^2+e_d^2+e_s^2) + \beta_1^{\rm EM}\,(e_a^2+e_b^2) + \beta_2^{\rm EM}\,(e_a-e_b)^2  \\
&+ \gamma_0^{\rm EM}\,(e_u^2\delta m_u+e_d^2\delta m_d+e_s^2\delta m_s)
+\gamma_1^{\rm EM}\,(e_a^2\delta \mu_a+e_b^2\delta \mu_b)  \\
&+ \gamma_2^{\rm EM}\, (e_a-e_b)^2\, (\delta \mu_a+\delta \mu_b)
+ \gamma_3^{\rm EM}\, (e_a^2-e_b^2)\, (\delta \mu_a-\delta \mu_b) \\
&+ \gamma_4^{\rm EM}\,(e_u^2+e_d^2+e_s^2)\,(\delta \mu_a +\delta \mu_b) 
+ \gamma_5^{\rm EM}\,(e_a+e_b)\,(e_u\delta m_u+e_d\delta m_d+e_s\delta m_s) \,. \end{split}
\label{mmasspq}
\end{equation}
For octet baryons Eq.~(\ref{bmass}) becomes
\begin{equation}
\begin{split}
M^2(aab) &= M_0^2 + \alpha_1\, (2\delta \mu_a +\delta \mu_b) +
\alpha_2\, (\delta \mu_a -\delta \mu_b)\\ 
&+ \beta_0^{\rm EM}\, (e_u^2+e_d^2+e_s^2) + \beta_1^{\rm
  EM}\,(2e_a^2+e_b^2) + \beta_2^{\rm EM}\,(e_a-e_b)^2  + \beta_3^{\rm
  EM}\,(e_a^2-e_b^2) \,.
\end{split}
\label{bmasspq}
\end{equation}
The coefficients $\alpha$, $\beta^{\rm EM}$ and $\gamma^{\rm EM}$ in
(\ref{mmasspq}) and (\ref{bmasspq}) are identical to those in
(\ref{mmass}) and (\ref{bmass}). This is to say that hadron mass splittings are unaffected by PQ at this order, as PQ moves (e.g.) all octet mesons and baryons by the same amount. We would have to expand to cubic terms to see PQ errors in the splittings~\cite{Bietenholz:2011qq}.
Hence, PQ calculations offer a computationally cheaper way of
obtaining them.  

In QCD + QED there is some ambiguity in the definition of the
symmetric point. 
The definition we have chosen is that the electrically neutral
pseudoscalar mesons have the same masses,
$M^2(u\bar{u})=M^2(d\bar{d})=M^2(s\bar{s})=M^2(d\bar{s}) = M^2(s\bar{d})=M^2(n\bar{n})
\,,$ where $n$ is a fictitious electrically neutral quark. 
As annihilation diagrams are neglected, different neutral mesons do
not mix. 
We denote the Wilson hopping parameter $\kappa$ (introduced in (\ref{faction}) below) marking the symmetric
point by $\bar{\kappa}_q$. 
We then have $\delta m_q = (m_q-\bar{m}) = 1/2\kappa_q^{\rm sea} - 1/2\bar{\kappa}_q$ and $\delta \mu_q = (\mu_q-\bar{m}) =
1/2\kappa_q^{\rm val} - 1/2\bar{\kappa}_q$, setting the lattice spacing $a = 1$.
It should be noted that even when all three quark masses are equal we
do not have full SU(3) symmetry. 
Because of their different charges, the $u$ quark is always
distinguishable from the $d$ and $s$ quark.

\section{Lattice matters}

The action we are using is  
\begin{equation} 
   S = S_G + S_A + S_{F}^u + S_{F}^d + S_{F}^s \,.
\label{action}
\end{equation}
Here $S_G$ is the tree-level Symanzik improved SU(3) gauge action with gauge coupling $\beta=6/g^2$,
and $S_A$ is the noncompact U(1) gauge action~\cite{Gockeler:1989wj,Gockeler:1991bu}
of the photon,
\begin{equation}
S_A=\frac{1}{2e^2} \,\sum_{x, \mu<\nu} \left(A_\mu(x) + A_\nu(x+\mu)
- A_\mu(x+\nu) - A_\nu(x)\right)^2 \,.
\end{equation}
We employ the nonperturbatively ${\cal O}(a)$ improved SLiNC fermion
action~\cite{Cundy:2009yy} for each quark flavor,
\begin{eqnarray}
   \tilde{S}_{F}^q = \sum_x \Big\{\frac{1}{2} 
                        \sum_{\mu}\big[\bar{q}(x)(\gamma_\mu - 1)
                          e^{-ie_q\,A_\mu(x)}\tilde{U}_\mu(x) q(x+\hat{\mu}) 
                  \!\!&-&\!\!
                        \bar{q}(x)(\gamma_\mu + 1) 
                          e^{ie_q\,A_\mu(x-\hat{\mu})}\tilde{U}^\dagger_\mu(x-\hat{\mu})
                          q(x-\hat{\mu})\big] 
                        \nonumber   \\
            + \frac{1}{2\kappa_q} \bar{q}(x)q(x) \!\!&-&\!\!
             \frac{1}{4} c_{SW} \sum_{\mu\nu}
                 \bar{q}(x)\sigma_{\mu\nu}F_{\mu\nu}(x)q(x)
                                        \Big\} \,.
\label{faction}
\end{eqnarray}
This action features single iterated mildly stout smeared QCD links with $\alpha=0.1$~\cite{Bietenholz:2011qq} and unsmeared QED links in the hopping terms, while the clover term contains unsmeared QCD links only. We keep the action deliberately local, as excessive smearing will lead to large autocorrelation times. Stout smearing is analytic, so a derivative can be taken, which makes the HMC force well defined. The clover coefficient has been computed nonperturbatively in QCD~\cite{Cundy:2009yy}. We presently neglect electromagnetic modifications to the clover term. This will leave us with corrections of $O(\alpha_{\rm EM}\,e_q^2\, a)$, which turn out to be no larger than the $O(a^2)$ corrections from QCD in our simulations. We check this later by comparing neutral meson masses with different quark charges $e_q$ (Fig.~\ref{neutral}). Adding an electromagnetic clover term with $c_{SW}^{\rm EM}=1$ would leave us with corrections of $O(\alpha_{\rm EM}\,e_q^2\, g^2 a)$ (to this order in $\alpha_{\rm EM}$), which is not a significant improvement, if at all. Simulations are performed using the HMC and RHMC~\cite{Clark:2006fx} algorithms. The gluon field and the EM field are updated sequentially.

In this study, we limit our calculations to a single value of the
strong coupling constant (lattice spacing) $\beta = 5.50$,
where we have our largest sample of dynamical QCD configurations
\cite{Bornyakov:2015eaa}. 
Furthermore, we restrict ourselves to simulations at the symmetric
point, $\delta m_u =\delta m_d = \delta m_s = 0$, which we define as
$X_\pi^2/X_N^2 = 0.126$, where $X_N^2 = \left(M_n^2 + M_p^2 +
  M_{\Sigma^-}^2 + M_{\Sigma^+}^2 + M_{\Xi^-}^2 +
  M_{\Xi^0}^2\right)/6$.
We may use either $X_\pi$ or $X_N$ to set the scale \cite{Bietenholz:2011qq}. 
After several tuning runs carried out on $24^3\times 48$ lattices we
arrived at the $\kappa$ values $\bar{\kappa}_u= 0.124362 \,, \ \bar{\kappa}_d = \bar{\kappa}_s = 0.121713$. At these $\kappa$ values, we study three different volumes, $24^3\times 48$, $32^3\times 64$ and $48^3\times 96$, with $O(2000)$ to
$O(500)$ trajectories. We like to add that simulations at the symmetric point already catch the essential features of the physical QCD + QED vacuum, as flavor singlet quantities vary slowly along the $\bar{m} = \mbox{constant}$ trajectory~\cite{Bietenholz:2011qq}.

On these ensembles we have computed PQ pseudoscalar meson and octet
baryon masses for a variety of quark masses ranging from $m_{PS}/m_N =
0.22$ to $0.5$, with $e_q=-1/3, 0$ and $+2/3$. 
This leads to about 40 pseudoscalar masses and 70 baryon masses per
ensemble. 
The baryons include several artificial states containing the
fictitious $n$ quark and charge 2 baryons with flavor structure
$uuu^\prime$.

\begin{figure}[b!]
\vspace*{-0.25cm}
   \begin{center} 
      \epsfig{file=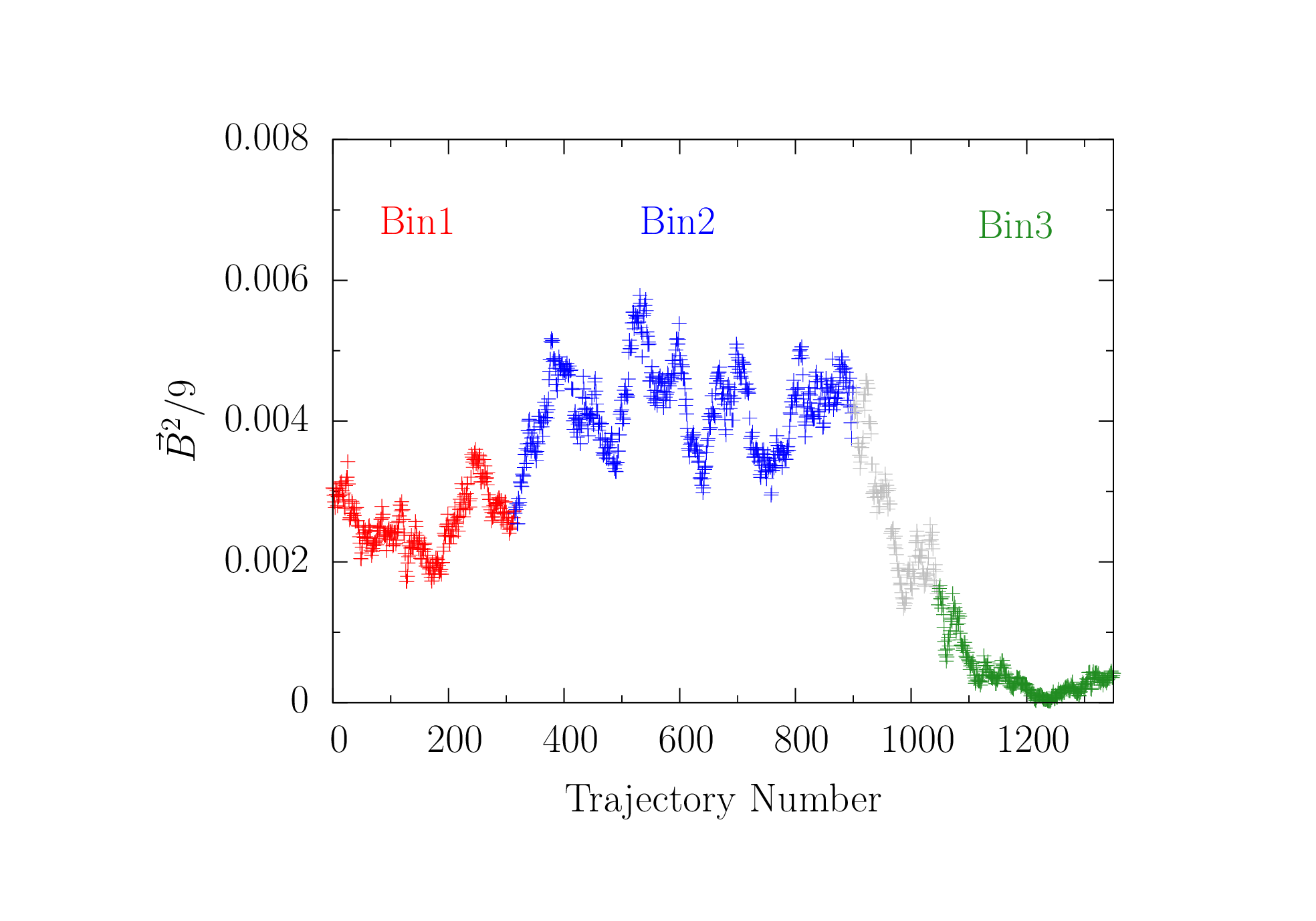,width=8.5cm,clip}\hspace*{-0.5cm}
      \epsfig{file=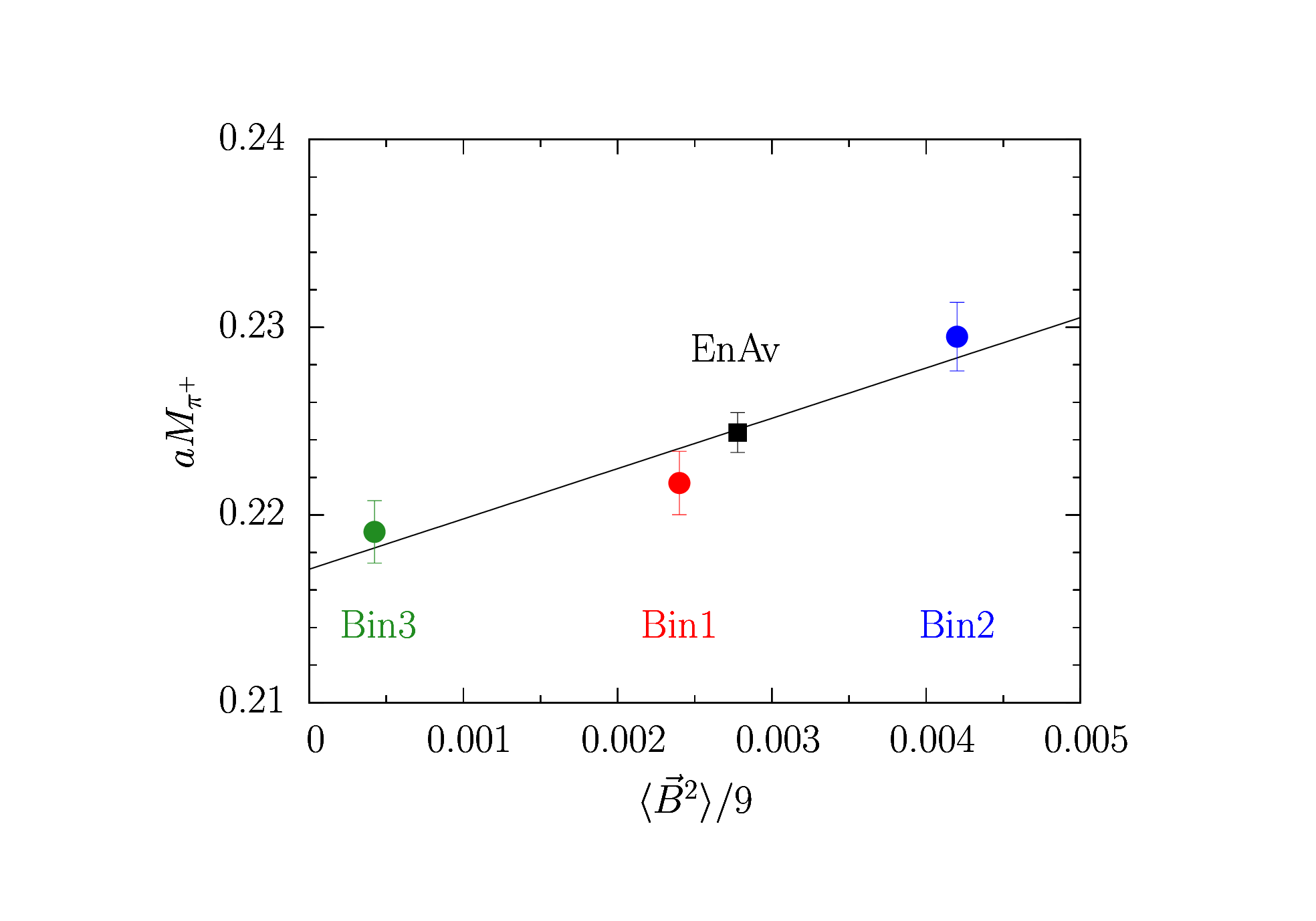,width=8.5cm,clip}
   \end{center} 
\vspace*{-1.0cm}
\caption{Left panel: The background field on the $24^3\times 48$ lattice divided into three bins of approximately constant $\vec{B}^2$. Right panel: The bin averaged energy of the charged pion at rest against the bin averaged $\langle\vec{B}^2\rangle$. The black square ($\blacksquare$) indicates the ensemble average (EnAv) of both $aM_{\pi^+}$ and $\langle\vec{B}^2\rangle$. The line is a one-parameter linear fit through the three mass points.}
\label{binplot}
\end{figure} 

The action (\ref{action}) is invariant under $\mathrm{U(1)}$ gauge transformations
\begin{equation}
A_\mu(x) \rightarrow A_\mu(x)+ \Delta_\mu\, \alpha(x) \,,\quad
q(x) \rightarrow e^{ie_q\alpha(x)}\,q(x) \,.
\end{equation}
However, this is not the case for propagators of charged particles,
which demands fixing the gauge, as in perturbation theory.
We choose the Landau gauge, which is defined by the condition
$\bar{\Delta}_\mu A_\mu(x) = 0$, where $\Delta_\mu$
($\bar{\Delta}_\mu$) is the forward (backward) lattice derivative.
The Landau gauge does not eliminate all gauge degrees of freedom, but
allows for shifts $\Delta_\mu \alpha(x)$ of the photon field with
$\Delta^2 \alpha(x) = 0$, where
$\Delta^2=\Delta_\mu\bar{\Delta}_\mu$~\cite{Gockeler:1991bu}.
To maintain (anti-)periodicity of the quark fields, $\alpha(x)$ must
be periodic up to a transformation of the form
\begin{displaymath}
e_q \alpha(x) = \sum_\mu \, \frac{2\pi}{L_\mu}\, n_\mu x_\mu \, ,
\quad n_\mu \in \mathbb{Z} \,, 
\end{displaymath}
where $L_\mu$ is the extent of the lattice in $\mu$ direction. 
This gauge field redundancy can be eliminated by adding multiples of
$2\pi/e_q L_\mu$ to $A_\mu(x)$, such that
\begin{equation}
- \frac{\pi}{|e_q| L_\mu} < B_\mu \leq \frac{\pi}{|e_q| L_\mu} \, ,
\quad B_\mu = \frac{1}{V} \sum_x A_\mu(x) \,.
\label{zmode}
\end{equation}
Taking $e_q=-1/3$ in (\ref{zmode}) serves both charges. The advantage of this
procedure is that it leaves the fermion determinant and Polyakov loops for all quark flavors unchanged. In other popular gauges~\cite{Borsanyi:2014jba} this is not the case, but results in a permanent Polyakov loop $\Pi_{x_\nu, \nu \neq \mu}\, L_\mu^{\rm EM}(x_\nu) = 1$, which we would not know how to correct for in a simple manner.

The constant background field can be factored out from the link matrices, configuration by configuration, and absorbed into the quark momenta by straightforward algebra.\footnote{Note that the transformation $\displaystyle q(x) \rightarrow e^{i\Delta x}\,q(x)\,,\; \bar{q}(x) \rightarrow \bar{q}(x)\, e^{-i\Delta x}$ amounts to a shift of quark momenta $\displaystyle p \rightarrow p+\Delta$.} This leaves us with photon propagators that are devoid of zero modes. In the presence of a constant background field $B_\mu$ the correlator of a single hadron $H$ thus  becomes~\cite{Gockeler:1991bu} 
\begin{equation}
\langle 0|H(t) \bar{H}(0)| 0\rangle \simeq |Z_H|^2 \,
e^{-\sqrt{M_H^2+\left(\vec{p}+e_H\vec{B}\right)^2} t}\,, 
\label{corr}
\end{equation}
where $M_H$, $\vec{p}$ and $e_H$ are mass, three-momentum and electric
charge of the hadron, respectively. 
This amounts to a shift of the rest energy of the charged hadrons,
$M_H \rightarrow \sqrt{M_H^2 + e_H^2\, \vec{B}^2} \simeq M_H + e_H^2\, \vec{B}^2/2M_H$.
We determine the ensemble average of $\vec{B^2}$ directly on each of
our three volumes. The result is $0.024$, $0.0079$ and $0.000095$ for the smallest to largest volumes, respectively. To extract masses, we remove the influence of the background field effect by subtracting the associated kinetic energy from the ensemble averaged lattice energy. To demonstrate the validity of this procedure, we have divided a subset of our $24^3\times 48$ ensemble into three bins of approximately constant background field in Fig.~\ref{binplot} and plot the corresponding lattice energies for each of these bins against the corresponding $\vec{B^2}$. It shows that both the energies of the individual bins as well as the ensemble averaged energy fall on a single straight line, in line with our subtraction method.    
On the $48^3\times 96$ lattice the effect of the background field is comparable to our statistical precision.
With the zero modes removed, we then can employ established methods, such as~\cite{Davoudi:2014qua}, to correct for the remaining electromagnetic finite size effects associated with the long-range tail of the photon field. Any residual effect of the background field will only act to modify the recoil energy of any charged hadron propagator within loops.

Our strategy is to simulate at an artificial coupling $e^2 = 1.25$,
and then interpolate between this point and pure QCD to the physical
fine structure constant $\alpha_{\rm EM} = 1/137$. This value is
chosen so that electromagnetic effects can be easily seen, but is
still small enough that they scale linearly in $e^2$ and we do not need to
consider higher order terms. Most importantly, $Z_3=0.94(3)$, obtained from the vacuum polarization. Furthermore, in Figs.~4 and 5 of \cite{Paul} we have plotted $1/\kappa_q^c\,,\, 1/\bar{\kappa}_q$ and the bare quark mass at the symmetric point, $1/2\bar{\kappa}_q - 1/2\kappa_q^c$, against $e_q^2$ for $e_q=-1/3, 0$ and $2/3$ and found that all three quantities lie on a straight line. In addition, we find that the coefficients $\alpha$ and $\alpha_1,\, \alpha_2$ in (\ref{mmasspq}) and (\ref{bmasspq}) agree to a good precision with the corresponding numbers in pure QCD~\cite{Bietenholz:2011qq}. This rules out significant higher order corrections in $e^2$.

\section{Results}

After the initial small volume tuning runs, it turns out that the
chosen $\kappa$ values do not quite satisfy our constraint
of equal neutral pseudoscalar meson masses. A more accurate estimate
can be determined from a fit to the pseudoscalar meson masses. On the
$48^3\times 96$ lattice we obtain
\begin{equation}
\bar{\kappa}_u= 0.124382 \,, \quad \bar{\kappa}_d = \bar{\kappa}_s = 0.121703, \quad \bar{\kappa}_n=0.120814 \,,
\end{equation}
which is only a small displacement from the underlying simulation
kappas. We shall expand about these $\kappa$ values in our subsequent
fits.

\begin{figure}[t!]
\vspace*{-2.5cm}
   \begin{center} 
      \epsfig{file=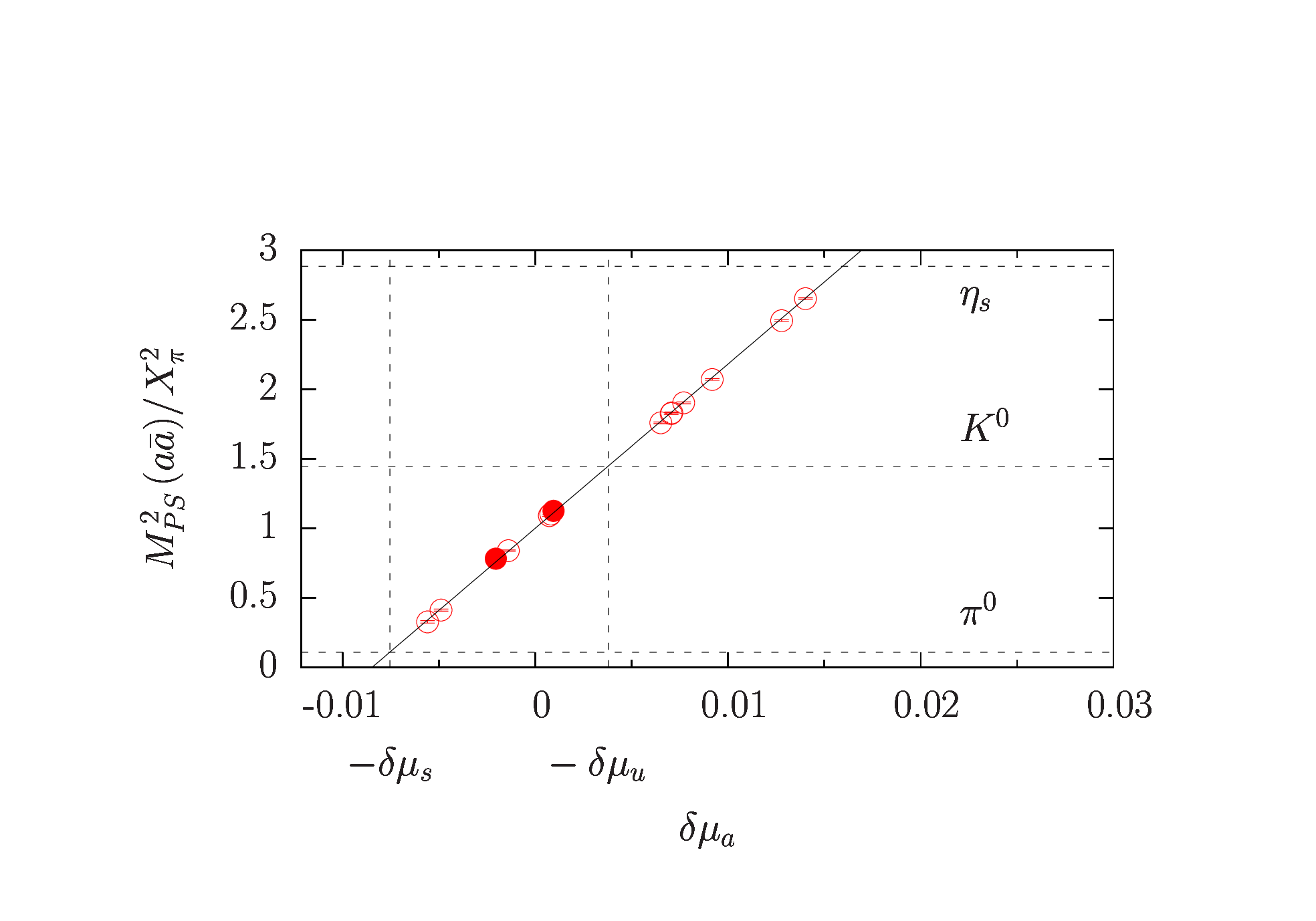,width=13cm,clip}
   \end{center} 
\vspace*{-1.5cm}
\caption{The neutral pseudoscalar meson masses $M_{PS}^2(a\bar{a})$ on the $48^3\times 96$ lattice as a function of $\delta \mu^D_a$ for quark charges $e_a=-1/3$ (blue), $0$ (green) and $2/3$ (red). Solid (open) symbols refer to unitary (PQ) masses. The horizontal lines display the physical $\pi^0, K^0$ and $\eta_s$ meson masses, the vertical lines indicate the physical $u$ and $s$ quark masses.}
\label{neutral}
\end{figure} 

\begin{figure}[t!]
\vspace*{-2.5cm}
   \begin{center} 
      \epsfig{file=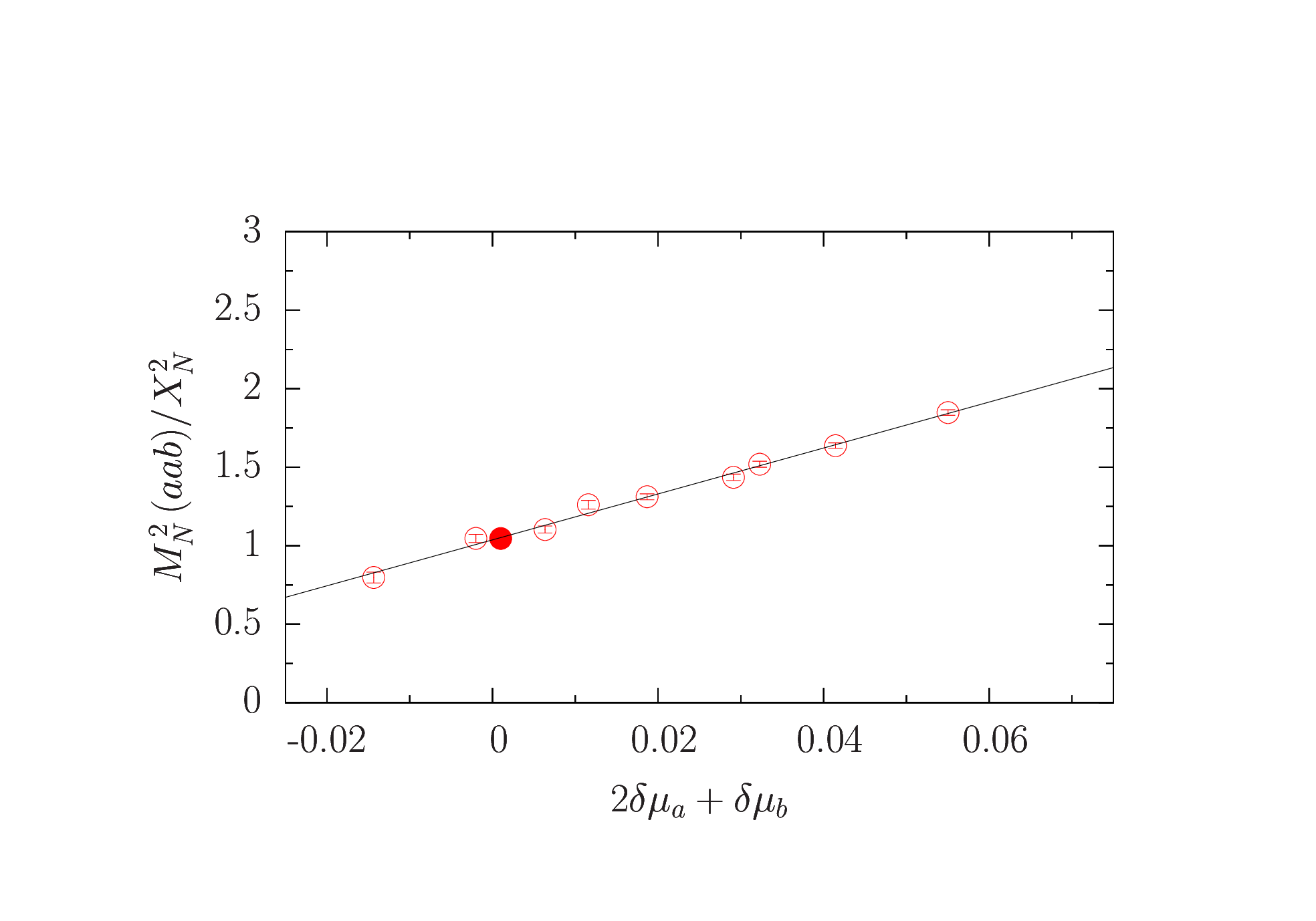,width=13cm,clip}
   \end{center} 
\vspace*{-1.5cm}
\caption{The charge $e_N=+1$ ($p$ and $\Sigma^+$) baryon masses $M_N^2(aab)$ on the $48^3\times 96$ lattice as a function of $2\delta \mu^D_a+\delta \mu^D_b$. Solid (open) symbols refer to unitary (PQ) masses.}
\label{chb}
\end{figure}

In contrast to QCD, equal meson masses at the symmetric point no longer mean equal bare quark masses. We renormalize the quark masses to remove this defect. We do so by absorbing the QED terms of the neutral pseudoscalar mesons into the quark self-energies. On our symmetric background, $\delta m_u = \delta m_d = \delta m_s = 0$, this is achieved by replacing $\delta \mu_q$ by the `Dashen' scheme mass~\cite{Paul}
\begin{equation}
\delta \mu_q^D = [1+(\gamma_1^{\rm EM}/\alpha)\,e_q^2]\, \delta \mu_q \,.
\label{Dmu}
\end{equation}
Substituting (\ref{Dmu}) into (\ref{mmasspq}), and absorbing $\beta_0^{\rm EM}$ into $M_0^2$ and $\gamma_4^{\rm EM}$ into $\alpha$, we obtain in the `Dashen' scheme 
\begin{equation}
\begin{split}
M^2(a\bar{b}) &= M_0^2 + \alpha\, (\delta \mu_a^D +\delta \mu_b^D) 
+ \beta_2^{\rm EM}\,(e_a-e_b)^2  \\
&+ \gamma_2^{\rm EM}\, (e_a-e_b)^2\, (\delta \mu_a^D+\delta \mu_b^D) + \gamma_3^{\rm EM}\, (e_a^2-e_b^2)\, (\delta \mu_a^D-\delta \mu_b^D) \,. 
\end{split}
\label{Dmmasspq}
\end{equation}
Note that since we choose the neutral pseudoscalar mesons to have the same mass, $\beta_1^{\rm EM}=0$ by definition.
We define the critical point, $\kappa_q^c$ for each flavor,  to be the point where the masses of the neutral pseudoscalar mesons vanish. It is then easily seen that the `Dashen' scheme quark masses are all equal at the symmetric point, $\bar{\mu}_q^D=M_0^2/2\alpha$, $q=u, d, s$ and $n$, see~\cite{Paul} for further details. It follows that the total electromagnetic contributions to the neutral pseudoscalar meson masses, $M_{\pi^0}$ and $M_{K^0}$, are zero. In Fig.~\ref{neutral} we show the neutral meson masses $M^2(a\bar{a})$ against $\delta \mu_a^D$. It is striking to see that the data fall perfectly on a straight line, which strongly supports our group-theoretical classification of SU(3) flavor symmetry breaking as well as octet (Gell-Mann--Okubo) type mass splitting. $O(\alpha_{\rm EM}\,e_a^2\, a)$ corrections would result in deviations from the straight line proportional to $e_a^2$, for which we see no evidence.  
To be consistent, we also expand the baryon masses in terms of the `Dashen' masses, 
\begin{equation}
\begin{split}
M^2(aab) &= M_0^2 + \alpha_1\, (2\delta \mu_a^D +\delta \mu_b^D) + \alpha_2\, (\delta \mu_a^D -\delta \mu_b^D)\\ 
&+ \beta_1^{\rm EM}\,(2e_a^2+e_b^2) + \beta_2^{\rm EM}\,(e_a-e_b)^2  + \beta_3^{\rm EM}\,(e_a^2-e_b^2) \,.
\end{split}
\label{Dbmasspq}
\end{equation}
In Fig.~\ref{chb} we show the charge $e_N=+1$ baryon masses $M^2(aab)$ against $2\delta \mu^D_a + \delta \mu^D_b$. Again, the data fall perfectly on a straight line, in accord with our flavor expansions.

\begin{figure}[h!]
\vspace*{-0.5cm}
   \begin{center} 
      \epsfig{file=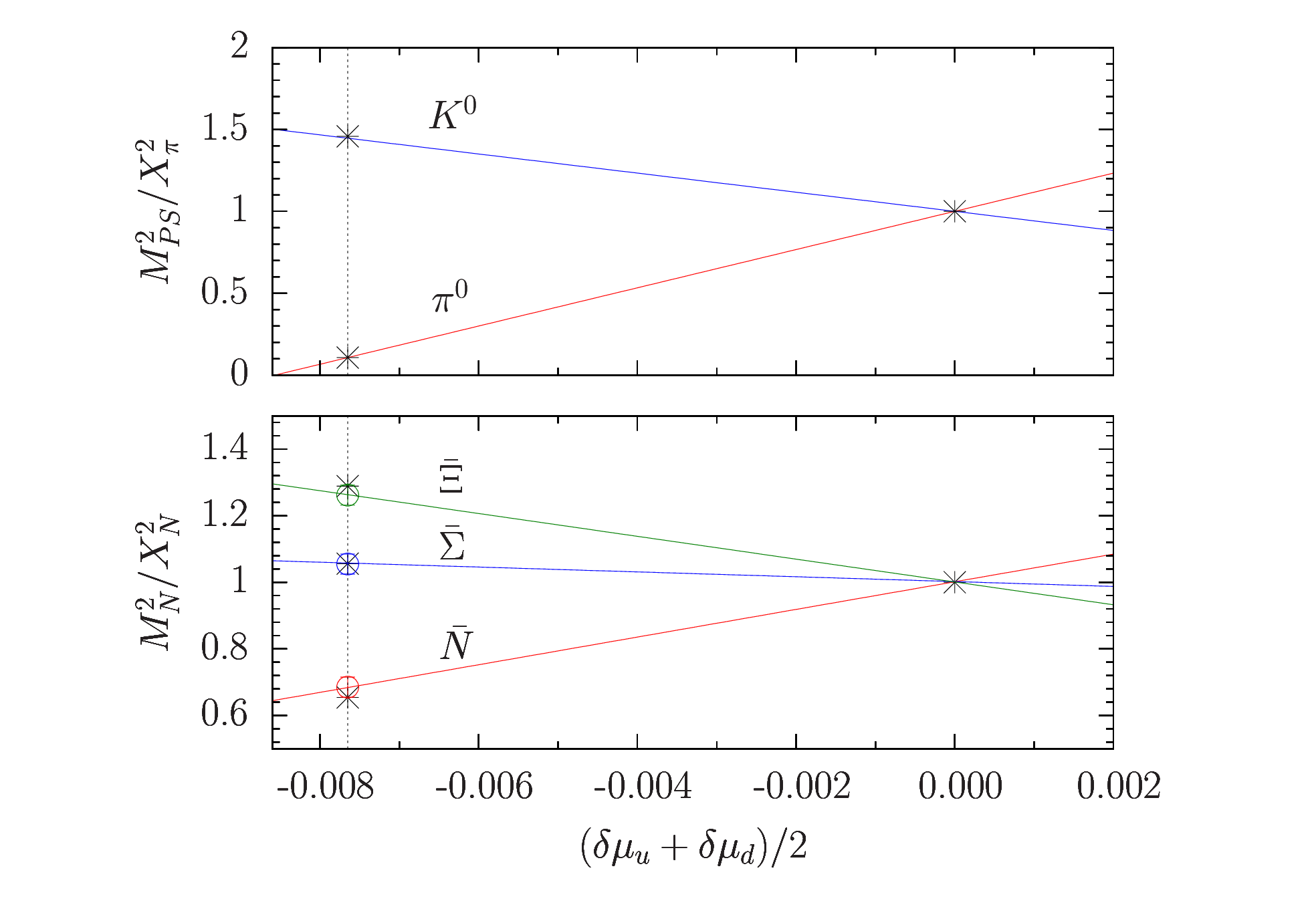,width=13cm,clip}
   \end{center} 
\vspace*{-1cm}
\caption{Fan plots of pseudoscalar meson (top) and baryon masses (bottom) on the $48^3\times 96$ lattice as a function of $\delta \mu_u + \delta \mu_d$, with $\delta \mu_u + \delta \mu_d + \delta \mu_s = 0$. The baryon masses are the averages of the isospin doublets.}
\label{fan}
\end{figure} 

For the total contribution of QCD + QED it does not matter which
scheme we use to define the quark masses, but for the individual
contributions of QCD and QED it will make a difference.  The fits of
(\ref{Dmmasspq}) and (\ref{Dbmasspq}) to the lattice data are quite
robust, giving $\chi^2/\mbox{dof} = 0.7 - 1.2$. To obtain physical
numbers, we extrapolate the coefficients $\beta_i^{\rm EM}$ and
$\gamma_i^{\rm EM}$ to $\alpha_{\rm EM}=1/137$ by scaling them with a
factor $\sim 10/137$. In our extrapolation to the physical point we
keep the sum of the quark masses constant. We choose $M_{\pi^0}^2$ and
$M_{K^0}^2-M_{K^+}^2+M_{\pi^+}^2-M_{\pi^0}^2$ to determine the
physical $\kappa$ values.
In Fig.~\ref{fan} we show the result of the fit to the meson and
baryon masses on the $48^3\times 96$ lattice. We obtain
$X_\pi^2/X_N^2=0.128(3)$, which is to be compared with the physical
value, $0.126$. This tells us that we have hit the symmetric point
with remarkable precision. Using $X_\pi$ to set the scale, the lattice
spacing turns out to be $a=0.068(2)\,\mbox{fm}$.
The figure also indicates that the baryon masses extrapolate nicely to
their experimental values, leaving little room for quadratic terms. 
Similarly good results are found on the $32^3\times 64$ lattice.
Having found the $\kappa$ values of the physical point and the point
where the `Dashen' scheme masses vanish (the critical point), we can
determine the quark masses. For the quark mass ratios we find on the
$48^3\times 96$ lattice in the `Dashen' scheme 

\begin{equation}
\frac{m_u}{m_d} = 0.52(2)\,, \quad \frac{m_s}{m_d} = 19.7(9)\,.
\label{eq:massratios}
\end{equation}
In \cite{Paul} we have shown how to switch between the `Dashen' and
$\overline{\rm MS}$ schemes.
Applying this, we find the ratio $m_u/m_d$ in the $\overline{\rm MS}$
scheme at $\mu^2=4\,{\rm GeV}^2$ decreases by less than a percent,
whereas $m_s/m_d$ remains a renormalization group invariant, even in
the presence of QED. Hence Eq.~(\ref{eq:massratios}) represents our
results in the \MSbar scheme at $\mu^2=4\gev^2$.

\begin{figure}[b!]
\vspace*{-0.5cm}
   \begin{center} 
      \epsfig{file=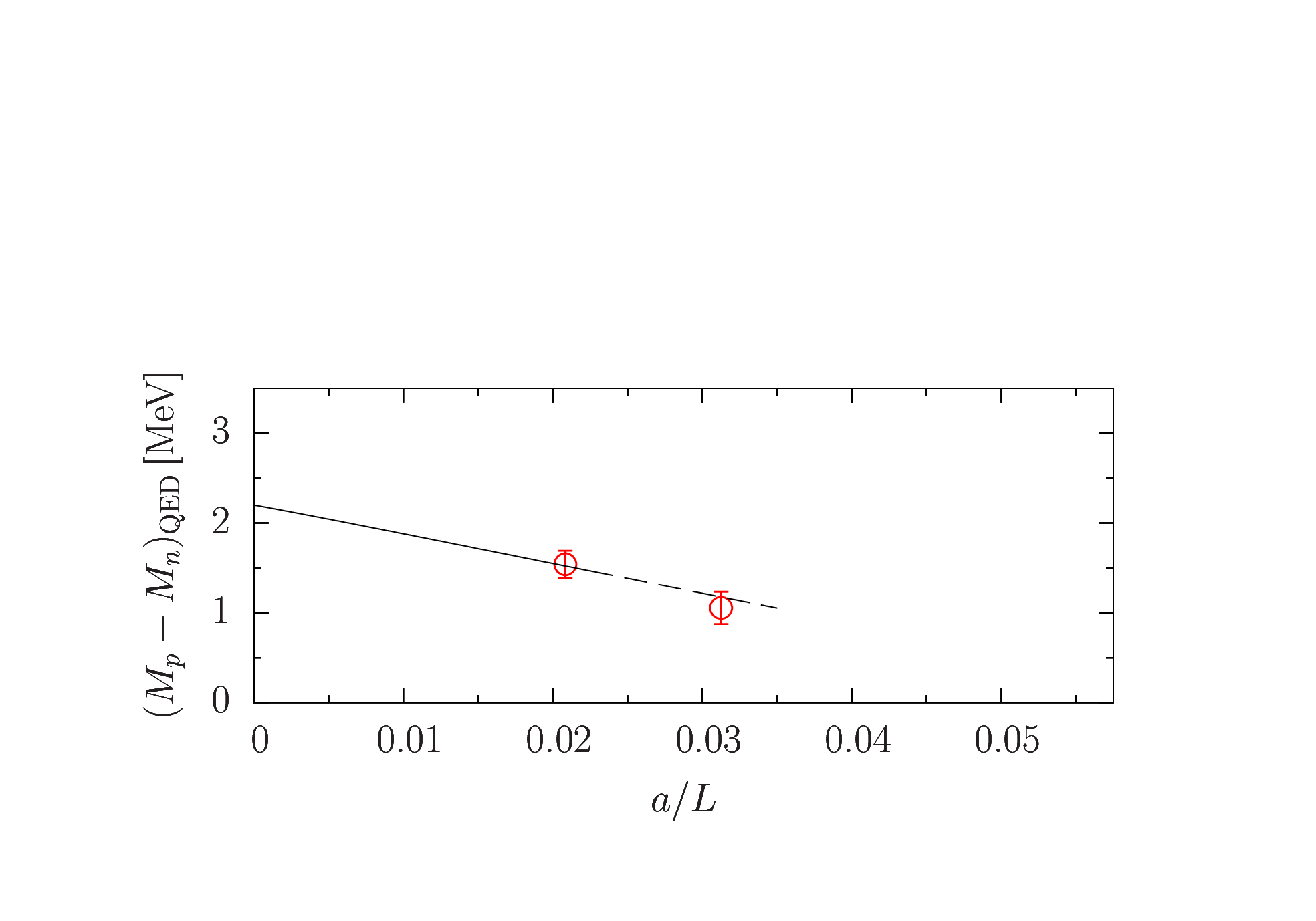,width=13cm,clip}
   \end{center} 
\vspace*{-1cm}
\caption{The QED contribution to the $p - n$ mass splitting on the $32^3\times 64$ and $48^3\times 96$ lattices compared with the prediction of~\cite{Davoudi:2014qua}.}
\label{test}
\end{figure} 

\begin{figure}[t!]
   \begin{center} 
      \epsfig{file=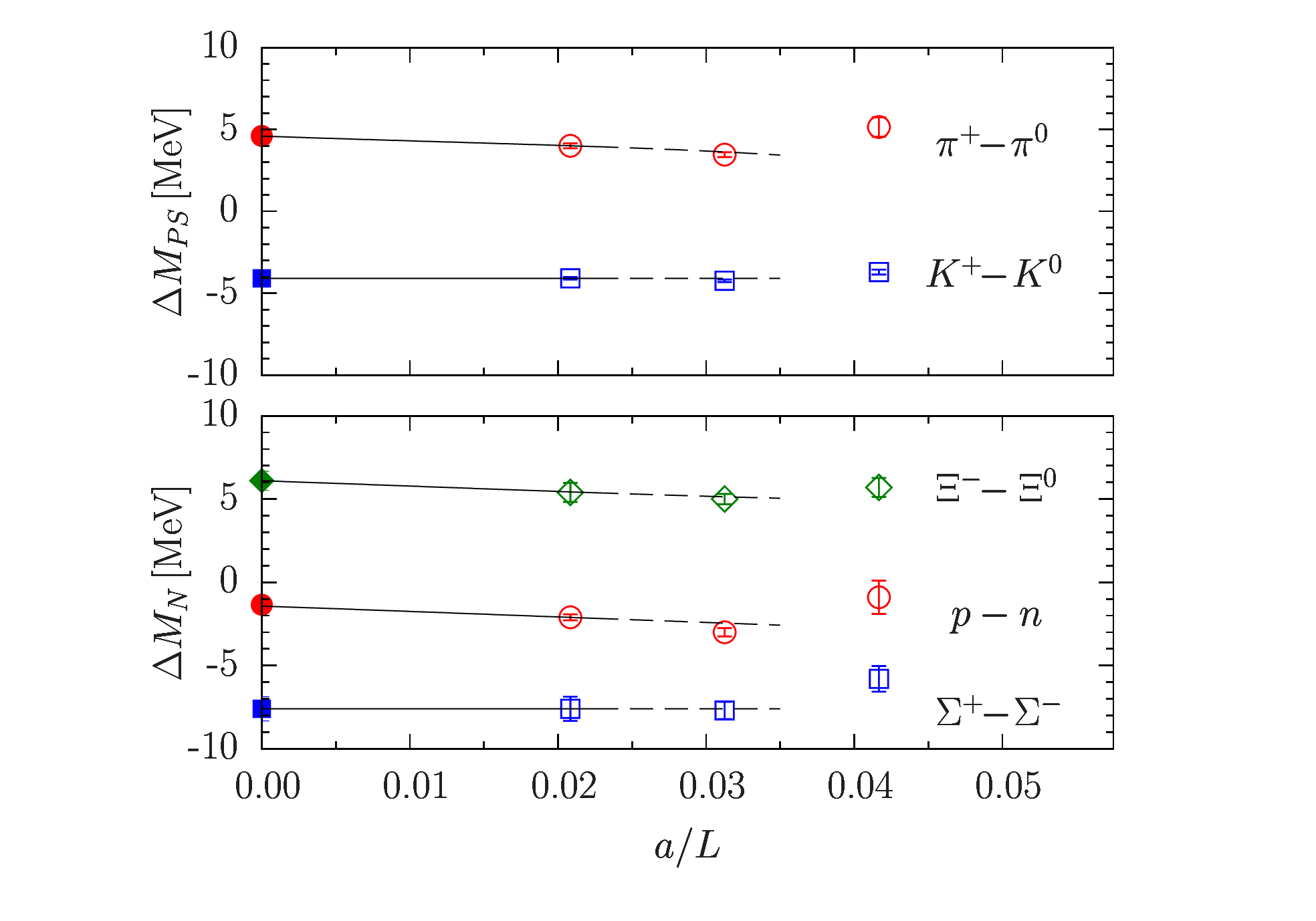,width=13cm,clip}
   \end{center} 
\vspace*{-1cm}
\caption{Mass splittings of pseudoscalar meson (top) and baryon masses (bottom) as a function of the spatial size of the lattice. The numbers on the largest volume have been extrapolated to infinite volume using~\cite{Davoudi:2014qua}.}
\label{ext}
\end{figure} 

\begin{figure}[b!]
\vspace*{-1.0cm}
   \begin{center} 
      \epsfig{file=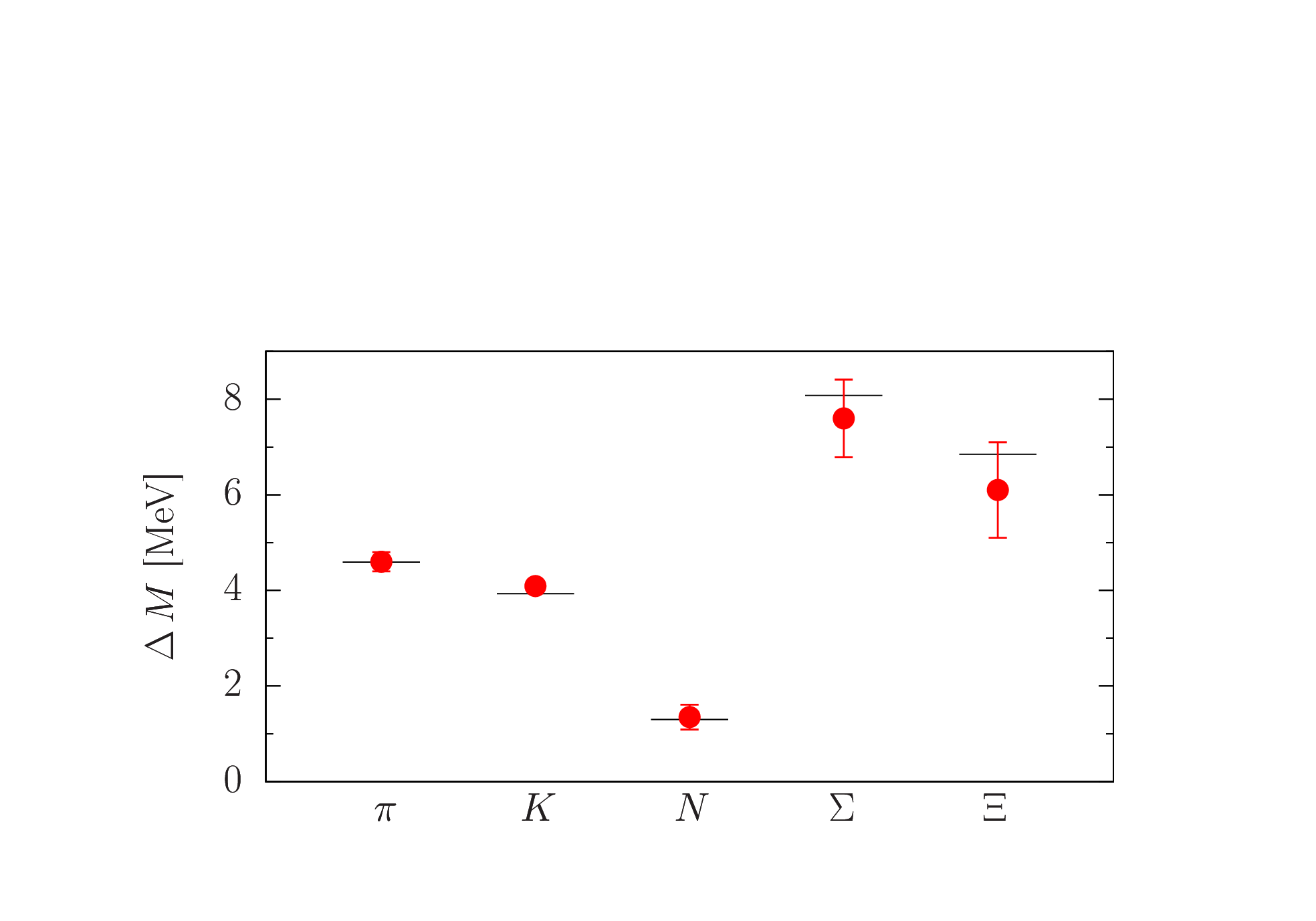,width=13cm,clip}
   \end{center} 
\vspace*{-1.0cm}
\caption{Mass splittings $\Delta M$ of octet pseudoscalar meson and baryon masses compared to experiment.}
\label{msplit}
\end{figure}

In this Letter we are primarily interested in the isospin splittings
of pseudoscalar meson and octet baryon masses. To get to our final numbers, we need to correct for finite size effects first. From QED we expect power-law corrections, due to the photon being massless, in addition to exponential corrections from QCD. We correct for QCD finite size effects by using the results of~\cite{AliKhan:2003ack,Colangelo:2005gd}, adapted to three flavors of PQ quarks. In case of the nucleon the corrections amount to approximately $1\%$ on the $48^3\times 96$ lattice and to $5\%$ on the $32^3\times 64$ lattice. Having successfully removed the zero modes, we can correct for the remaining QED effects by employing the mass shift formulae of effective field theory (EFT)~\cite{Davoudi:2014qua}. We test this in Fig.~\ref{test}, where we compare the QED contribution to the $p - n$ mass splitting, $(M_p-M_n)_{\rm QED}$, with the prediction of~\cite{Davoudi:2014qua} on our two largest volumes. We find good agreement between the data and the analytic expression, indicating that QED finite size effects are well accounted for by EFT. In Fig.~\ref{ext} we present our QCD + QED results for the isospin splittings of mesons and baryons as a function of lattice size. 
The curves represent the predictions of~\cite{Davoudi:2014qua}. They have been drawn through the points on the $48^3\times 96$ lattice. We find good agreement between the curves and our points on the two largest lattices, while the data on the $24^3\times 48$ lattice (with $L \approx 1.6\,\mbox{fm}$) appear to lie outside the range of validity of the expansion. We consider the extrapolation of the $48^3\times 96$ lattice points to $a/L = 0$ by~\cite{Davoudi:2014qua} our best estimate of the infinite volume result. We compare this result with a fit to the points on the two largest lattices. The differences are taken as an estimate of systematic error.
In Table~\ref{tabsplit} we list our final results for the mass splittings in
the infinite volume, for the total and the QED contribution
separately. Following \cite{Paul}, we find the QED contributions in the `Dashen' scheme and the \MSbar scheme at $\mu^2=4\gev^2$ to differ by less than
a percent. As a result, the QED contributions in Table~\ref{tabsplit} also represent our results in the \MSbar scheme at $\mu^2=4\gev^2$.
The traditional way of expressing the electromagnetic contributions is
through $\Delta_\pi=M_{\pi^+}^2-M_{\pi^0}^2$ and the $\epsilon$
parameter,
\begin{equation}
(M_{K^+}^2-M_{K^0}^2)_{\rm QED}-M_{\pi^+}^2+M_{\pi^0}^2=\epsilon\,\Delta_\pi \,.
\end{equation}
On the $48^3\times 96$ lattice we find $\epsilon=0.49(5)$, 
which translated to $\overline{\rm MS}$ gives \cite{Paul}
\begin{equation}
\epsilon = 0.50(6) \,.
\label{eps}
\end{equation}
\begin{table}[!t]
\begin{center}
\begin{tabular}{c|c|c|c|c}
  $\Delta M$            & \ QCD + QED      & \ QED         & QCD~\cite{Horsley:2012fw} & Experiment\\[0.1em] \hline
& & & & \\[-1.0em]
$M_{\pi^+}-M_{\pi^0}$     &                & $\phantom{-}4.60(20)\phantom{(00)}$  & & $4.59$\\[0.2em]
$M_{K^0}-M_{K^+}$         & $4.09(10)\phantom{(00)}$ & $-1.66(6)\phantom{(00)0}$  &          & $3.93$\\[0.2em]
$M_n-M_p$                & $1.35(18)(8)\phantom{0}$ & $-2.20(28)(10)$    & $3.51(31)$      & $1.30$\\[0.2em]
$M_{\Sigma^-}-M_{\Sigma^+}$ & $7.60(73)(8)\phantom{0}$  & $-0.63(8)(6)\phantom{00}$    & $9.07(47)$        & $8.08$\\[0.2em]
$M_{\Xi^-}-M_{\Xi^0}$      & $6.10(55)(45)$ & $\phantom{-}1.26(16)(13)$ & $5.58(31)$ & $6.85$\\ 
\end{tabular}
\end{center}
\vspace*{-0.25cm}
\caption{Mass splittings in the infinite volume, in units of MeV. The first error is the statistical error from the extrapolation of the points on the $48^3\times 96$ lattice. The second error (if any) is a systematic error estimated from the fit to both the $48^3\times 96$ and $32^3\times 64$ volumes.
The QCD + QED and QED results are compared with previous results from pure QCD~\cite{Horsley:2012fw} and the experimental numbers.}
\label{tabsplit}
\end{table}
This result is well within the range quoted by FLAG
\cite{Aoki:2013ldr}, albeit with significantly reduced uncertainty.
We now can compare the baryon mass splittings of this calculation with our recent results from pure QCD~\cite{Horsley:2012fw}. The QCD numbers are quoted in the fourth column of Table~\ref{tabsplit}. They have been brought in line with our new value of $\epsilon$ (\ref{eps}). Both sets of results are found to be largely consistent. It is worth emphasizing that the QED and pure QCD contributions to the nucleon mass splitting sum up nicely to the total QCD + QED contribution, which is encouraging. Finally, in the last column of Table~\ref{tabsplit} we quote the experimental mass splittings. We observe good agreement for both octet pseudoscalar mesons and octet baryons. Since we have not yet computed the QCD
contribution to the $\pi^0$ mass from $\pi^0$--$\eta$ mixing, arising from quark-line disconnected
diagrams, we only quote the QED contribution to the
$M_{\pi^+}-M_{\pi^0}$ mass difference.
It is worth noting that phenomenological estimates for the disconnected contribution are of the order of 0.1 MeV \cite{Gasser:1982ap}, which
is within the precision of our present calculation. Figure~\ref{msplit} summarizes our results. 

Both the total QCD + QED mass splittings as well as the QED contributions satisfy the Coleman-Glashow relation~\cite{Coleman:1961jn} by construction. So do the experimental values, which once again supports our group-theoretical approach and truncation (\ref{Dbmasspq}). The QED contribution to the $n - p$ mass splitting in the `Dashen' and \MSbar schemes turns out to be somewhat larger (in absolute terms) than the numbers derived from the Cottingham
formula~\cite{WalkerLoud:2012bg}. It should be noted though that the individual estimates~\cite{WalkerLoud:2012bg} cover a wide range of values. To accommodate the lower numbers from the Cottingham formula, the result of pure QCD~\cite{Horsley:2012fw} (fourth column of Table~\ref{tabsplit}) would have to be smaller by a factor up to two as well.
Our QED result is also larger than the recently reported lattice number in~\cite{Borsanyi:2014jba}. In our approach the QED and QCD separation is defined within the meson sector. In contrast, \cite{Borsanyi:2014jba} chose the QED part of the $\Sigma^+ - \Sigma^-$ mass difference to be zero, for which we identify a clear nonzero signal. This would be the case if $(2/3)\,\beta_1^{\rm EM}+\beta_2^{\rm EM}+(1/3)\,\beta_3^{\rm EM}=0$ in our mass expansion (\ref{Dbmasspq}). A fit to our data with this constraint gives $(M_n-M_p)_{\rm QED}=-1.71(28)(10) \,\mbox{MeV}$ in the `Dashen' scheme. While this result is largely compatible with the analysis of Walker-Loud, Carlson and Miller~\cite{WalkerLoud:2012bg}, $(M_n-M_p)_{\rm QED} = -1.30(50)\, \mbox{MeV}$, it illustrates quite clearly that the QED part of the $n-p$ mass difference depends sensitively on how electromagnetic and strong contributions are separated. While our results do not support higher order terms in the quark mass expansion, it may be possible that one source of the discrepancy could
be related to nonlinearities in the chiral behavior of the
electromagnetic self energy \cite{Thomas:2014dxa} that are not being
captured by the Taylor expansion. 

\begin{figure}[t!]
\vspace*{-1.25cm}
   \begin{center} 
      \hspace*{-1.5cm}\epsfig{file=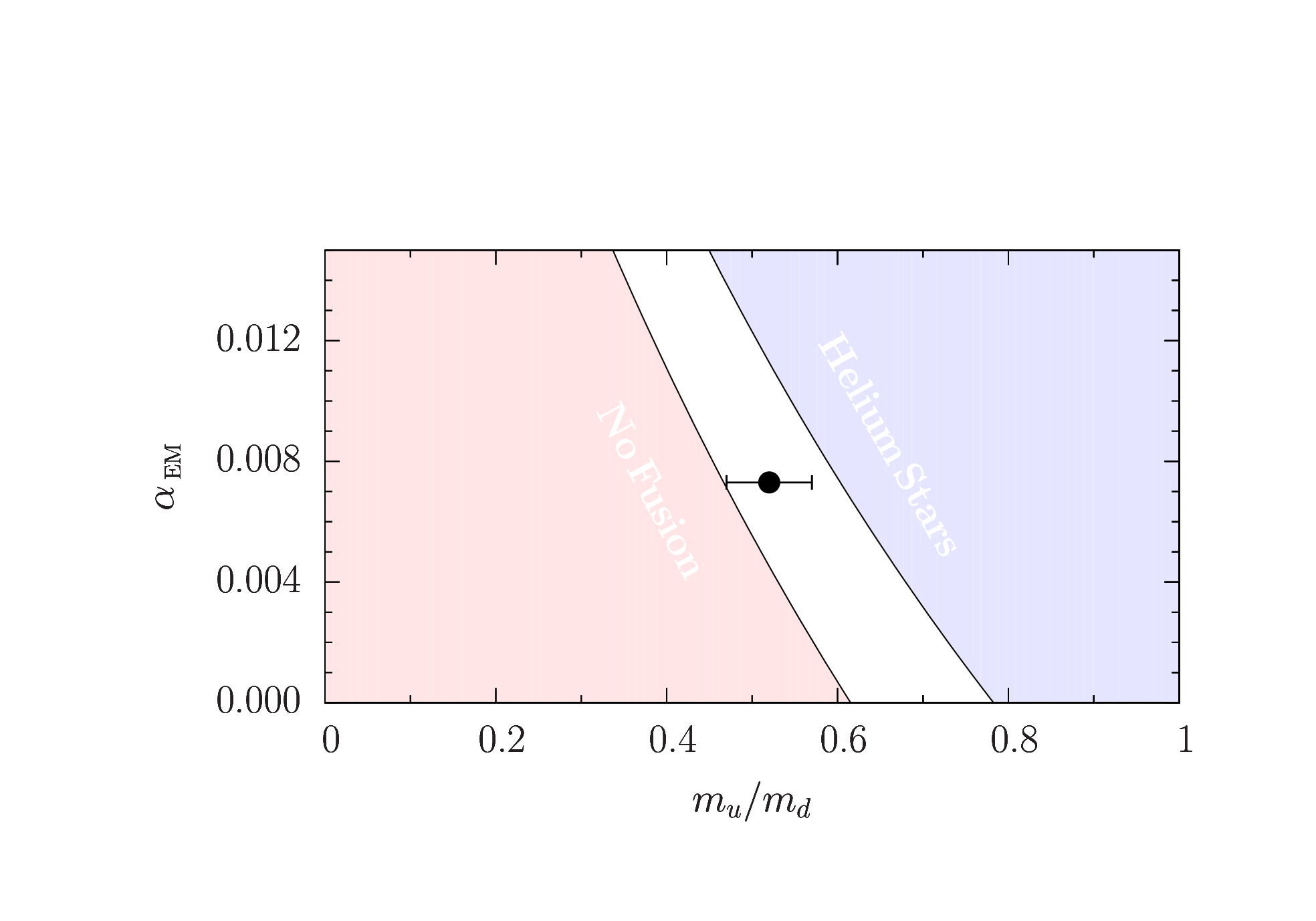,width=13cm,clip}
   \end{center} 
\vspace*{-1cm}
\caption{The allowed ratio of quark masses $m_u/m_d$ for a range of against $\alpha_{\rm EM}$. The solid circle is our result in the \MSbar scheme. The region of no fusion is to the left, the region where all hydrogen is converted to helium stars is to the right.}
\label{com}
\end{figure}

As discussed in the introduction, the existence of the Universe as we
know it is highly sensitive to the magnitude of the $n - p$ mass
difference. Having an analytic expression for the mass of neutron and proton, Eq.~(\ref{Dbmasspq}), we can express the allowed region in terms of the fundamental parameters $m_u, m_d$ and $\alpha_{\rm EM}$, as shown in Fig.~\ref{com}. Not shown are the bounds on $\alpha_{\rm EM}$ from the stability of atoms~\cite{Lieb:2001ec}. It turns out that both $\alpha_{\rm EM}$ and the ratio of light quark masses $m_u/m_d$ are finely tuned. At the physical fine structure constant the ratio is restricted to a narrow region around $m_u/m_d=0.5$.    

\section{Conclusion and outlook}

We have outlined a program to systematically investigate the flavor
structure of hadrons in a full QCD + QED lattice simulation. 
By treating the valence quark masses
differently to those in the sea allows for a
range of valence quark masses and charges to be explored and
significantly enhances our ability to accurately constrain the fit
parameters in our flavor-breaking expansions. As a result, we have
successfully computed the isospin splittings of pseudoscalar meson and
octet baryon masses. 
By using our recently introduced `Dashen' scheme as an intermediate
step \cite{Paul}, we are able to quote the first lattice results for
the QED contribution to the $n-p$ mass splitting in the
${\overline{\rm MS}}$ scheme.

The calculations have been done at lattice spacing $a=0.068\,\mbox{fm}$. At this lattice spacing discretization errors are expected to be less than $2\%$~\cite{Bornyakov:2015eaa}, which are well below our present statistical and systematic errors. To reduce the errors and gain full control over the infinite volume extrapolation, simulations on $64^3\times 128$ lattices and larger will have to be done. To further constrain our fits, and test for potential $\delta m_q$ effects, we have started dynamical $1+1+1$ flavor simulations along the $\bar{m}=\mbox{const}$ line, with $\delta m_u \neq \delta m_d \neq \delta m_s \neq 0$ and sea quark masses approaching the physical point. Finally, future simulations will also naturally be required on lattices with different lattice spacings to allow for a continuum extrapolation.

\section*{Acknowledgment}

This work has been partly supported by DFG through grant SCHI 422/10-1 (HP) and  the Australian Research Council through grants DP140103067 (RDY and JMZ), FT120100821 (RDY) and FT100100005 (JMZ). The numerical calculations were carried out on the BlueGeneQs at FZ J\"ulich and EPCC Edinburgh (using DIRAC 2), the Cray XC30 at HLRN (Berlin and Hannover) and on the NCI National Facility (Canberra).

\end{document}